\documentclass[reprint,showpacs,showkeys,amsmath,amssymb,superscriptaddress,aps,prl]{revtex4-1}

\usepackage{graphicx}
\usepackage{float}
\usepackage{epstopdf}
\begin{document}
\preprint{Stinshoff {\it et al.}, Mn$_{1.5}$FeV$_{0.5}$Al}

\title{Completely compensated ferrimagnetism and sublattice spin crossing
       in the half-metallic Heusler compound Mn$_{1.5}$FeV$_{0.5}$Al. }

\author{Rolf Stinshoff}
\affiliation{Max Planck Institute for Chemical Physics of Solids, 01187 Dresden, Germany}

\author{Ajaya K. Nayak}
\affiliation{Max Planck Institute for Chemical Physics of Solids, 01187 Dresden, Germany}
\affiliation{Max Planck Institute of Microstructure Physics, 06120 Halle, Germany}

\author{Gerhard H. Fecher}
\affiliation{Max Planck Institute for Chemical Physics of Solids, 01187 Dresden, Germany}

\author{Benjamin Balke}
\affiliation{Institut f{\"u}r Anorganische und Analytische Chemie, Johannes Gutenberg - Universit{\"a}t, 55099 Mainz, Germany}

\author{Siham Ouardi}
\affiliation{Max Planck Institute for Chemical Physics of Solids, 01187 Dresden, Germany}

\author{Yurii Skourski}
\affiliation{Dresden High Magnetic Field Laboratory (HLD), 01328 Dresden, Germany}

\author{Tetsuya Nakamura}
\affiliation{Japan Synchrotron Radiation Research Institute, SPring-8, Hyogo 679-5198, Japan}

\author{Claudia Felser}
\affiliation{Max Planck Institute for Chemical Physics of Solids, 01187 Dresden, Germany}

\date{\today}

\begin{abstract}

The Slater--Pauling rule states that $L2_1$ Heusler compounds with 24 valence
electrons do never exhibit a total spin magnetic moment. In case of strongly
localized magnetic moments at one of the atoms (here Mn) they will exhibit a
fully compensated half-metallic ferrimagnetic state instead, in particular, when
symmetry does not allow for antiferromagnetic order. With aid of magnetic and
anomalous Hall effect measurements it is experimentally demonstrated that
Mn$_{1.5}$V$_{0.5}$FeAl follows such a scenario. The ferrimagnetic state is
tuned by the composition. A small residual magnetization, that arises due to a
slight mismatch of the magnetic moments in the different sublattices results in a
pronounced change of the temperature dependence of the ferrimagnet. A
compensation point is confirmed by observation of magnetic reversal and sign
change of the anomalous Hall effect. Theoretical models are presented that
correlate the electronic structure and the compensation mechanisms of the
different half-metallic ferrimagnetic states in the Mn-V-Fe-Al Heusler system.

\end{abstract}


\maketitle

Half-metallic ferromagnets are promising candidates for application in
spintronics because they exhibit 100\% spin polarization. They are metallic in
one spin direction and semiconducting in the other~\cite{Groot83,Groot08}.
However, ferromagnets produce a large dipole field that hinders the device
performance. For example, the dipolar magnetic anisotropy becomes very large for
in-plane magnetic systems leading to large switching fields. For this reason,
there is a great interest on zero magnetic moment spintronics, as such systems
do not produce dipole fields and are extremely stable against external
magnetic fields~\cite{Soh11,Park11,Wang12,Marti14}. The concept of half-metallic
antiferromagnetism was introduced by van~Leuken and de~Groot~\cite{Groot95}. It
turns out, however, that symmetry does not allow half-metallic antiferromagnets
and the materials are half-metallic compensated ferrimagnets~\cite{Wurmehl06}.
Recently, Hu~\cite{Hu12} presented a theoretical work on possible half-metallic
antiferromagnets for spintronic applications. However, the identical electronic
structure of both spin directions makes most of the conventional
antiferromagnets unable to carry a spin polarized current.

Heusler materials are well known for their tunable magnetic structure due to the
presence of one or more magnetic sublattices. Depending on the constituting
elements or crystal structure, ferromagnetic, ferrimagnetic, antiferromagnetic,
or canted spin structures may be realised~\cite{Webster88,Ziebeck01,Kuebler00,Graf11,Felser16}.
In particular, the Heusler compounds with $L2_1$ or $C1_b$ structure are well
known for their half-metallic behaviour~\cite{Groot83}. These materials follow the
Slater--Pauling rule~\cite{Slater36,Pauling38} related to the half-metallicity~\cite{Galanakis02,Fecher06,Kandpal07}. 
According to this rule the spin magnetic moment ($m$) in cubic Heusler compounds
with $L2_1$ structure is defined by $m=N_v-24$, where $N_v$ is the accumulated
number of valence electrons. As a direct consequence, Heusler compounds with
$N_v=24$ never exhibit a macroscopic magnetic moment.

In certain cases, however, the $DO_3$ or $L2_1$ Heusler compounds with 24
valence electrons are able to exhibit a fully compensated half-metallic
behaviour~\cite{Wurmehl06}. In that concept, the Slater--Pauling rule is combined
with the K{\"u}bler rule~\cite{Kubler83}. The latter states that Mn on the
octahedrally coordinated position ($4b$) in Heusler compounds tends to a high,
localised magnetic moment. This moment has to be completely compensated by the
magnetic moments of the remaining atoms to satisfy the Slater--Pauling rule.

Although there are several theoretical predictions, most of the suggested
materials either do not exist or appear only in a different crystal
structure~\cite{Groot95}. Recently it was demonstrated that a compensated
ferrimagnetic state may be realized in the tetragonal Mn-Pt-Ga
system~\cite{Nayak15}. However, it is known that Heusler materials with tetragonal
distortion do not show half-metallicity. Kurt {\it et al}~\cite{Kurt14,Betto15}
have shown that a compensated magnetic state with considerable 
spin polarization may be achieved in a cubic thin film of Mn$_2$Ru$_x$Ga with
composition falling between $C1_b$ and $L2_1$ Heusler compounds. Despite several
attempts by different research groups there is no experimental evidence of a
compensated magnetic structure in the classical 24 valence electron based cubic
Heusler compounds. In the present work it is shown by experiments and calculations
that the Heusler compound Mn$_{1.5}$V$_{0.5}$FeAl with
$L2_1$ structure exhibits a completely compensated magnetic state. Further, the presence of
a temperature and composition dependent sublattice spin compensation is
demonstrated in the investigated system, while keeping the half-metallicity.


Polycrystalline ingots of Mn$_{1.5}$V$_{0.5}$FeAl were prepared by arc melting.
The composition and structure of the samples was determined by energy dispersive
X-ray spectroscopy (EDX) and X-ray powder diffraction (XRD).
Low field magnetic measurements were carried out by means of a vibrating sample
magnetometer (MPMS~3, Quantum Design). Pulsed, high magnetic field experiments
were performed at the Dresden High Magnetic Field Laboratory. The transport
measurements were carried out utilising a physical property measurement system
(PPMS, Quantum Design). X-ray magnetic circular dichroism (XMCD) investigations
were performed at beamline BL25SU of SPring-8. The electronic structure was
calculated in the local spin density approximation. The selfconsistent
electronic structure calculations were carried out using the spin polarized
fully relativistic Korringa--Kohn--Rostocker method (SPRKKR) provided by Ebert
{\it et al}~\cite{Ebert99,Ebert11}.

\begin{figure}[htb]
\centering
\includegraphics[width=\columnwidth]{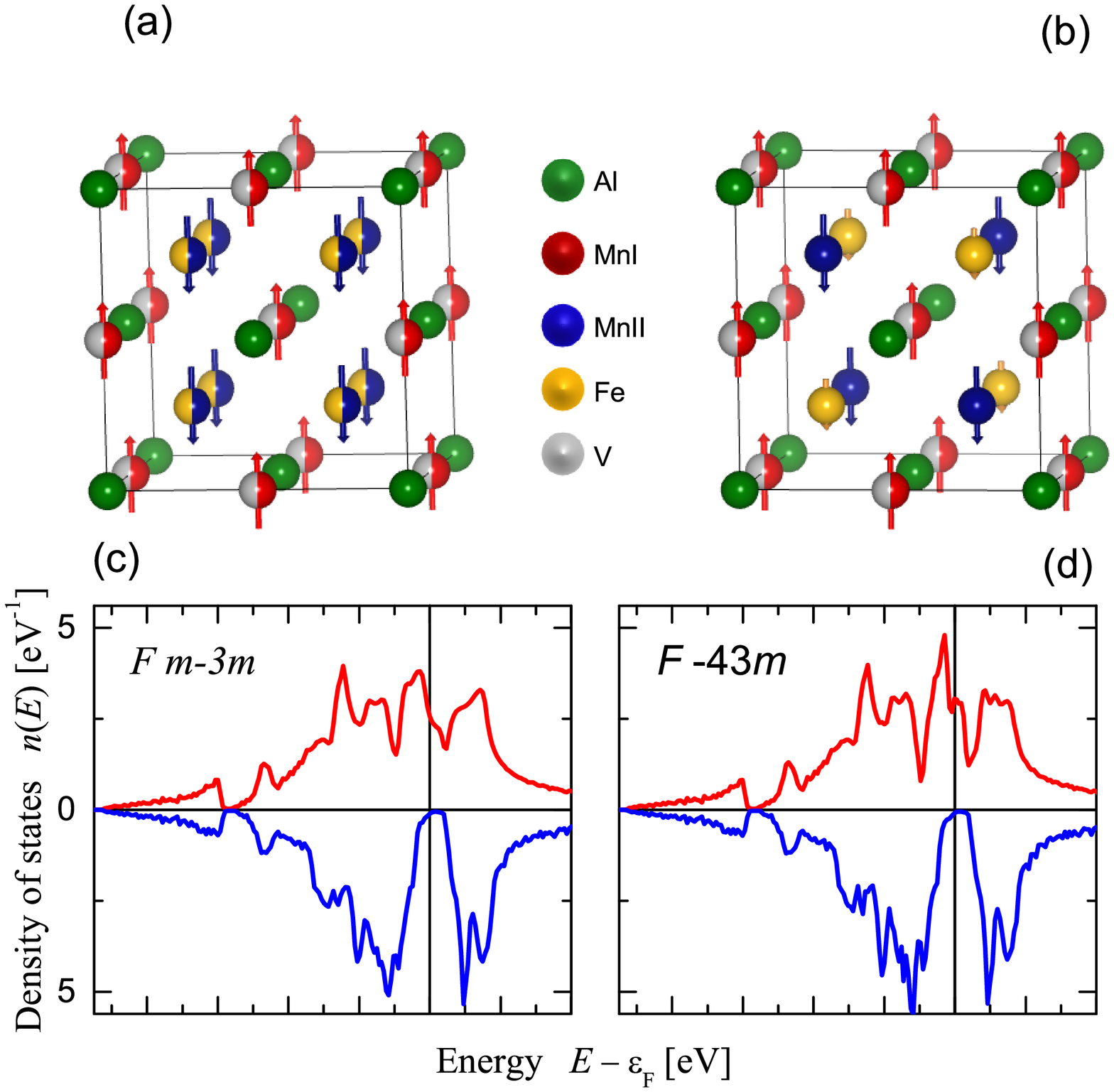}
\caption{(Color online) Crystalline and electronic structure of Mn$_{1.5}$V$_{0.5}$FeAl \newline
         (a) The $L2_1$ type cubic Heusler structure with space group $F\:m\overline{3}m$ (225).
         (b) The $X$ type inverse cubic Heusler structure with space group $F$-43$m$ (216).
         Different atoms are represented by balls with different colours, shown in between the two structures.
         The spin resolved density of states is shown for $F\:m\overline{3}m$ in (c)
				 and for $F\:\overline{4}3m$ in (d).}
\label{fig:1_struct}
\end{figure}

The XRD analysis indicates that Mn$_{1.5}$V$_{0.5}$FeAl crystallizes in a cubic
Heusler structure with a lattice parameter of $a=5.83$~{\AA}. The two most
energetically favoured crystal structures are shown in
Figure~\ref{fig:1_struct}(a) and~(b) together with their magnetic order. In the
regular $L2_1$ cubic Heusler structure with space group $F\:m\overline{3}m$
(225), the Al atoms occupy the $4a$ position, the $4b$ position is equally
occupied by V and Mn atoms and  a statistical distribution of the Mn and Fe
atoms at $8c$ is expected. In a less probable situation, an ordering of the Mn
and Fe atoms can split the $8c$ position to $4c$ and $4d$, as shown in
Figure~\ref{fig:1_struct}(b). In order to determine the density of states (DOS)
of Mn$_{1.5}$V$_{0.5}$FeAl, the calculations were performed using SPRKKR with
coherent potential approximation (CPA) to account for the random occupation of
the sites and for chemical disorder. Comparing the total energies
at the same lattice parameter, one finds that the energy of the $L2_1$ structure
with space group $F\:m\overline{3}m$ (Figure~\ref{fig:1_struct}(a)) is 0.5 meV
lower compared to the $X$ structure with space group $F\:\overline{4}3m$
(Figure~\ref{fig:1_struct}(b)). The electronic structure reveals clearly the
half-metallic character of Mn$_{1.5}$V$_{0.5}$FeAl for both structure types with
chemical disorder. The gap in the minority DOS is defined by the states of the
Mn atoms located on the $8c$ and the Fe atoms located on the $8c$ or $4c$
positions. This coincides with previous calculation for various Heusler
compounds, as in most cases, the gap is dominated by the states arising from the
atom on the $8c$ site~\cite{Kandpal07}.

\begin{table}[htb]
\caption{Site specific magnetic moments in Mn$_{1.5}$FeV$_{0.5}$Al.\\
         The calculations were carried out by means of SPRKKR - CPA using the $L2_1$ structure
         ($F\:m\overline{3}m$, 225) or the $X$-type structure
         ($F\:\overline{4}3m$, 216). All magnetic moments
         are given in $\mu_B$. The total moments are given per primitive cell. The site
         specific spin $m_s$ and orbital $m_l$ magnetic moments are given per atom.
         Note the rounding, the induced moment at Al is $<0.006\:\mu_B$. }
\begin{center}
\begin{ruledtabular}
   \begin{tabular}{l ccc ccc}
                & \multicolumn{3}{c}{225} & \multicolumn{3}{c}{216} \\
      Atom      & Site   & $m_s$ & $m_l$  &  Site  & $m_s$ & $m_l$  \\
      \hline
      Mn        & ($8c$) &  1.40  &  0.03  & ($4d$) &  1.38  &  0.03  \\
      Fe        & ($8c$) &  0.28  &  0.02  & ($4c$) &  0.35  &  0.03  \\
      Mn        & ($4b$) & -2.79  & -0.01  & ($4b$) & -2.86  & -0.01  \\
       V        & ($4b$) & -0.55  &  0.01  & ($4b$) & -0.60  &  0.01  \\
      Al        & ($4a$) & -0.01  & -0.00  & ($4a$) & -0.01  & -0.00  \\
      \hline
      $m^{s,l}_{tot}$ &  &  0.003 &  0.046 &        & 0.000  &  0.058 \\
      $m_{tot}$ & & \multicolumn{2}{c}{0.05} & & \multicolumn{2}{c}{0.06} \\
   \end{tabular}
\end{ruledtabular}
\end{center}
\label{tab:table1}
\end{table}

The calculated magnetic moments of Mn$_{1.5}$FeV$_{0.5}$Al are listed in
Table~\ref{tab:table1}. A summation of the site specific magnetic moments yields
a zero total moment, as expected for a completely compensated ferrimagnet. The
signs of the calculated magnetic moments were supported by XMCD measurements. XMCD spectra were calculated for the
$L2_1$ structure with Mn on $8c$ mixed with Fe and on $4b$ mixed with V. The two
different Mn atoms cause a zero-crossing with pronounced maximum and minimum at
the $L_3$ edge, that is also clearly revealed in the measured spectra. Although
the site specific moment of the Mn atoms could not be determined due to overlap
of the lines from the Mn atoms at two different sites, the total sum moment and
moments obtained for Fe and V from the XMCD measurements well matched with the
theoretical values.

\begin{figure}[htb]
\centering
\includegraphics[width=\columnwidth]{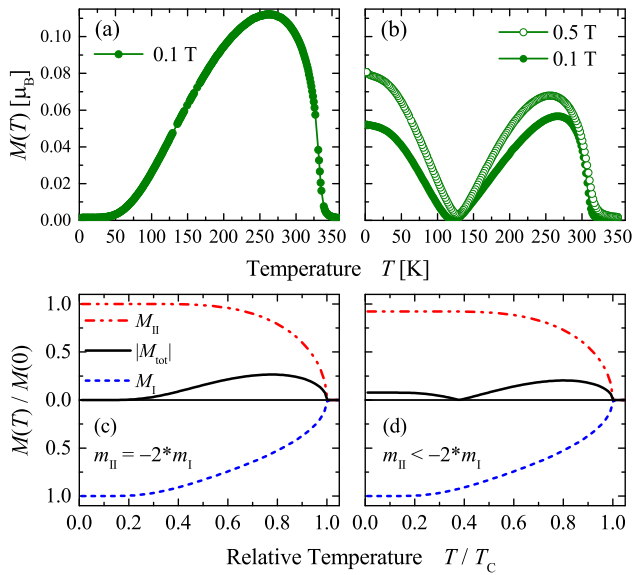}
\caption{(Color online) Magnetisation of Mn$_{1.5}$FeV$_{0.5}$Al. \newline
         (a) shows the temperature dependence of the
             magnetization $M(T)$, measured for a completely compensated sample.
         (b) shows the temperature dependence of
             the magnetization measured for an overcompensated sample in different fields. 
         The theoretical behaviour of a two sublattice ferrimagnet in the 
         molecular field approximation is shown in (c) and (d):
         (c) shows the magnetization for a completely compensated ferrimagnet with $| m_{II} | = |2m_{I} |$,
         (d) shows the case with $| m_{II} | < | 2m_{I} |$.
              $m_{i}$ are the average magnetic moments of the atoms on the $i^{\rm th}$
              sublattice at $T=0$.}
\label{fig:2_magn}
\end{figure}

Figure~\ref{fig:2_magn}a shows the temperature dependence of the magnetisation
$M(T)$ for a completely compensated sample. As expected, the magnetization
vanishes at 0~K as is typical for a completely compensated ferrimagnet. 
The magnetisation stays close to Zero up to about 50~K. The Curie temperature
appears at about 335~K.

$M(T)$ curves measured for a slightly overcompensated sample in different
induction fields are shown in Figure~\ref{fig:2_magn}b. From the $M(T)$ curves
measured in an induction field of 0.1~T a Curie temperature ($T_{\rm C}$) of
about 308~K is determined. By decreasing the temperature the magnetization first
completely reduces to zero at 127~K and then increases again by lowering the
temperature below 127~K. This type of magnetic behaviour indicates the presence
of a compensation point of the ferrimagnetic order. The completely compensated
behaviour is very sensitive to the composition of the sample as will be shown
next.

The temperature dependence of the total and the sublattice magnetic moments were
simulated using a molecular field model for a two sublattice ferrimagnet. In
particular the equations introduced by Stearn~\cite{Stearns68} for binary compounds
(Fe$_3$Al, Fe$_3$Si) with Heusler type structure were used. This model may come
close to the $L2_{1}$ structure with space group 225. It is assumed that the
magnetic moment $m_I$ of the atoms in sublattice I is smaller (half) but that
twice as many atoms are occupying lattice I. That is, lattice I describes the
$8c$ site, whereas lattice II corresponds to the $4b$ sites with higher magnetic
moments $m_{II}$ but only half as many atoms ($n_I/n_{II}=2$) compared to
lattice I. The completely compensated ferrimagnet appears when
$n_Im_I=n_{II}m_{II}$. The exchange integrals $J_{ij}$ should be largest for
interactions between the atoms in sites I and II. Further the exchange integrals
between atoms of type I should be much smaller compared to the atoms of type
II, with the latter being close to those between type I and II atoms. In
particular it was assumed that $J_I/J_{I-II}=1/2$ and $J_{II}/J_{I-II}=2/3$. The
results are shown in Figures~\ref{fig:2_magn}c and~\ref{fig:2_magn}d that
compare the completely compensated case with a slightly overcompensated case.
Figure~\ref{fig:2_magn}c describes a ferrimagnet where the compensation point
appears at $T=0$ and the magnetization stays nearly Zero up to about
$T/T_C\approx1/5$. For $m_{II}<|2m_{I}|$, a compensation point appears
(Figure~\ref{fig:2_magn}d). The latter is classified as a  N{\'e}el N-type
ferrimagnet~\cite{Neel71}.

\begin{figure}[htb]
\centering
\includegraphics[width=\columnwidth]{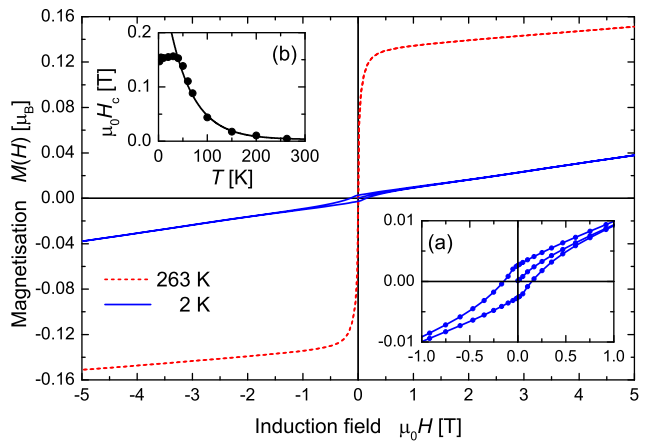}
\caption{(Color online) Field dependent magnetisation of Mn$_{1.5}$FeV$_{0.5}$Al. \newline
          Shown is the magnetisation $M(H)$ at 2~K and 263~K.
          Inset (a) shows the  low temperature behaviour on an enlarged scale. Inset (b) shows the temperature variation of the coercive field.
          }
\label{fig:3_M-H}
\end{figure}

Figure~\ref{fig:3_M-H} shows the field dependence of the magnetisation for two
different temperatures, at the maximum of the magnetisation (263~K) and close to
Zero (2~K). At high temperatures the material appears very soft but has a small
remanence and coercive field at low temperature. The inset~(b) shows that the
coercive field is constant below 50~K where the magnetisation vanishes (compare
Figure~\ref{fig:2_magn}a). Above this critical temperature, the magnetisation
softens with increasing temperature. The appearance of a coercive field is 
a typical effect at the compensation point~\cite{Webb88}. In certain
cases it is assumed to diverge at the compensation point, whereas it clearly 
saturates below the critical temperature in the completely compensated 
half-metallic ferrimagnet.

So far the occurrence and some magnetic properties of a completely compensated
half-metallic ferrimagnet is demonstrated. The compensation phenomenon is better
studied, however, in the slightly overcompensated sample with a compensation
point at a finite temperature. The remaining part is thus devoted to this case.

\begin{figure}[htb]
\centering
\includegraphics[width=\columnwidth]{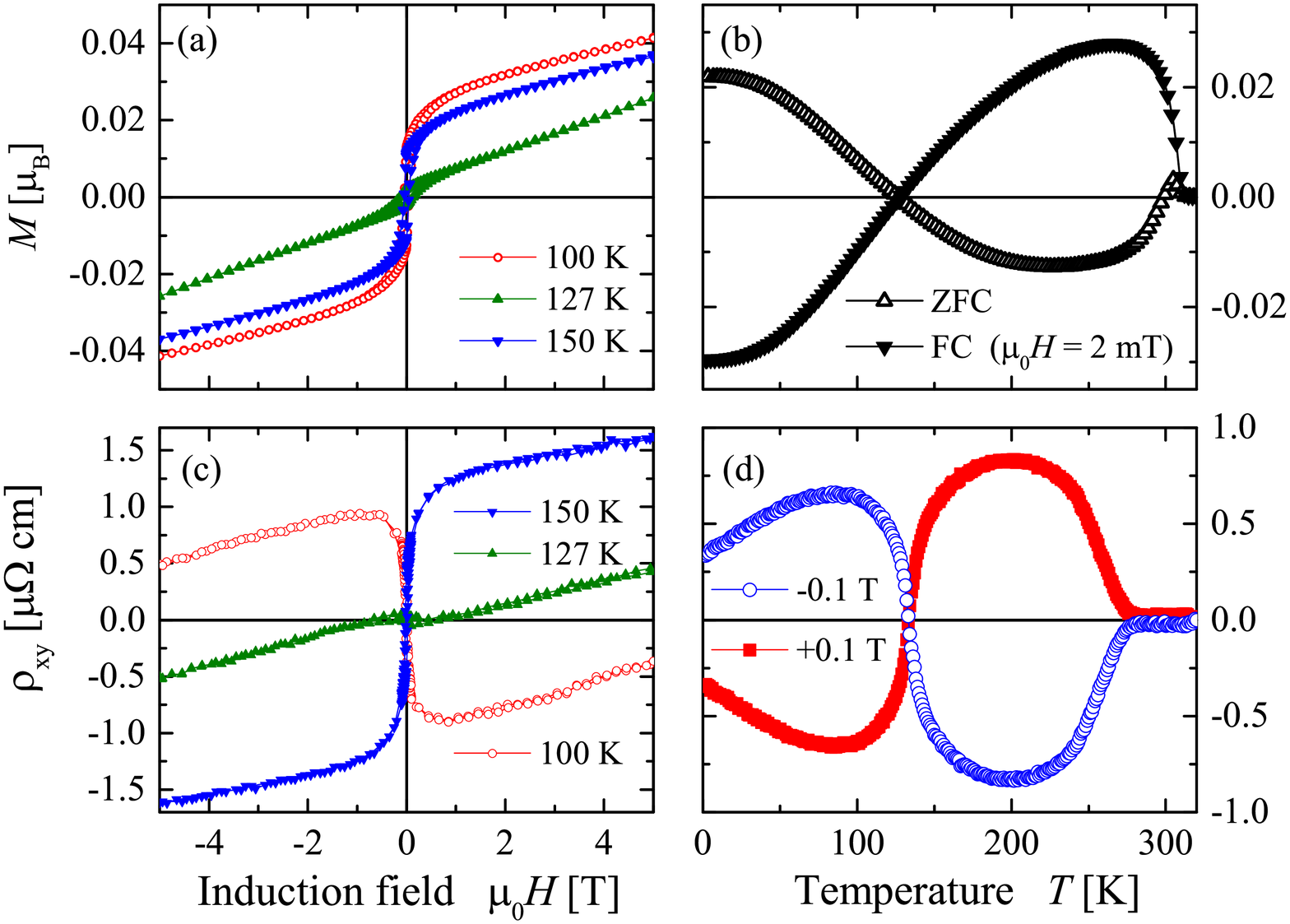}
\caption{(Color online) Magnetic and transport properties of overcompensated
                        Mn$_{1.5}$FeV$_{0.5}$Al. \newline
         (a) Isothermal magnetization loops, $M(H)$, at different temperatures.
         (b) ZFC and FC $M(T)$ curves measured in a small field of 2~mT.
         (c) Field dependence of the Hall effect measured at different temperatures.
         (d) Temperature dependence of Hall resistivity measured in $\pm$0.1~T.}
\label{fig:4_transp}
\end{figure}

The $M(H)$ loops measured at different temperatures demonstrate the compensation
phenomenon (Figure~\ref{fig:4_transp}a) in the overcompensated sample. A nearly
linear hysteresis loop with almost zero spontaneous magnetization is found in
the vicinity of the compensation temperature (127~K). The $M(H)$ loops measured
for temperatures below and above the compensation point exhibit a soft magnetic
behaviour. The most important point is that both $M(T)$ and $M(H)$ measurements
hint on a saturation magnetization that is less than 0.1$\mu_{\rm B}$ away from
the compensation point. This suggests that the sample virtually exhibits a
nearly compensated magnetic state over the full temperature range.

It is seen from Figure~\ref{fig:2_magn}b that the minimum at the compensation
point shifts slightly with increasing induction field. For a deeper
understanding of this effect we have measured Zero field cooled (ZFC) and field
cooled (FC) $M(T)$ curves in a very small field of 2~mT
(Figure~\ref{fig:4_transp}b). In this case the FC curve, which shows a positive
magnetization at higher temperature, crosses the temperature axis at 127~K to
give a negative magnetization at low temperatures. The ZFC curve follows an
exactly opposite behaviour to that of the FC curve. The zero-crossing of the
magnetization clearly indicates a sublattice magnetic compensation at 127~K.
Similar magnetic reversal at the compensation point has been observed in systems
with spin-orbital compensation~\cite{Taylor02}. Pulsed magnetic field measurements
at 1.5~K and at the compensation point (127~K) show a linear magnetic response
with fields up to 55~T without any spin-flop transition. This clearly indicates a strong exchange coupling
between the different magnetic sublattices in Mn$_{1.5}$V$_{0.5}$FeAl.

The magnetic measurements shown in Figures~\ref{fig:2_magn}(a) and~(b) only give
an indication of a sublattice spin crossing at the compensation point. Anomalous
Hall effect (AHE) measurements have been performed at different temperatures, to
allow for a direct observation of the spin crossing across the compensation
point (see Figure~\ref{fig:4_transp}). The AHE measured at 50~K and 100~K shows
a negative sign, i.e, negative (positive) value in positive (negative) field. At
the compensation point the AHE becomes virtually zero. Above the compensation
point for $T=150$~K and 200~K, a positive anomalous Hall effect is observed. The
change in sign of the AHE can be seen in the temperature dependence of the AHE measured
in a field of $\pm$0.1~T (Figure~\ref{fig:4_transp}d). The AHE changes from a
negative (positive) maximum around 100~K to a positive (negative) maximum at
200~K when measured in a field of 0.1~T (-0.1~T). The two curves cross the zero
line at about 130~K. Since the AHE is an intrinsic property of ferro- and
ferrimagnets, a small uncompensated moment below and above the compensation
point will result in a non-vanishing AHE. The most important point is that the
AHE changes its sign, which clearly indicates the change of the sublattice
magnetic structure across the compensation point. The AHE is an intrinsic
manifestation of a Berry curvature, that changes due to the change of the
sublattice magnetic moment from spin-up to spin-down, resulting in a change of
the sign across the compensation point.


In conclusion, the existence of a completely compensated ferrimagnetic state in
the half-metallic $L2_1$ cubic Heusler compound Mn$_{1.5}$V$_{0.5}$FeAl has been
experimentally demonstrated. Although there have been several theoretical works
regarding realization of a fully compensated magnetic state in the $L2_1$ cubic
Heusler compounds with 24 valence electrons, no successful experimental attempt
has been made until now. This work also establishes the existence of a
temperature dependent sublattice spin crossing in half-metallic ferrimagnets. The
compensation temperature can be varied by an intentional variation of the
stoichiometry. Recently, it has been demonstrated that antiferromagnets may be
utilized as a principal component in spintronic devices, especially in tunnel
magnetoresistance based devices. The present half-metallic compensated
ferrimagnet adds the advantage of nearly 100\% spin polarization, which is
extremely important for spintronics.


\begin{acknowledgments}

The authors thank N. Demitri for assistance during the XRD experiment at
ELETTRA. This work is funded by the Deutsche Forschungs Gemeinschaft (project
1.3-A in research unit 1464 {\it ASPIMATT}) and by the ERC Advanced Grant
(291472) {\it Idea Heusler}. The experiments at the High Magnetic Field
Laboratory Dresden (HLD) were supported by Euro-MagNET II under the European
Union contract 228043. X-ray diffraction measurements were performed at beamline
XRD1 of the ELETTRA Synchrotron (Trieste, Italy) under proposal number 20145509.
Synchrotron based HAXPES and XMCD measurements were performed at BL47XU and
BL25SU of SPring-8 with approval of JASRI, Proposal Nos. 2008A0017 and
2008A1606, respectively.

\end{acknowledgments}


\end{document}